\documentclass{article}

\usepackage{arxiv}

\usepackage[utf8]{inputenc} % allow utf-8 input
\usepackage[T1]{fontenc}    % use 8-bit T1 fonts
\usepackage{hyperref}       % hyperlinks
\usepackage{url}            % simple URL typesetting
\usepackage{booktabs}       % professional-quality tables
\usepackage{amsfonts}       % blackboard math symbols
\usepackage{nicefrac}       % compact symbols for 1/2, etc.
\usepackage{microtype}      % microtypography
\usepackage{graphicx}
\usepackage[square,sort,comma,numbers]{natbib}
\usepackage{doi}
\graphicspath{ {./images/} }

\title{On Identifying Points of Semantic Shift \\ Across Domains}

\author{
    \href{https://orcid.org/0000-0002-4075-0768}{\includegraphics[scale=0.06]{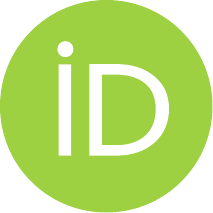}
    \hspace{1mm}Hyung Wook Choi}\\
    College of Computing and Informatics \\
    Drexel University \\
    Philadelphia, PA, 19104\\
    \texttt{hc685@drexel.edu} \\
       \And
    \href{https://orcid.org/0000-0002-0236-7389}{\includegraphics[scale=0.06]{orcid.pdf}\hspace{1mm}Mat Kelly} \\
    College of Computing and Informatics \\
    Drexel University \\
    Philadelphia, PA, 19104\\
    \texttt{mkelly@drexel.edu} \\
}

\begin{document}
\maketitle
\begin{abstract}
The semantics used for particular terms in an academic field organically evolve over time. Tracking this evolution through inspection of published literature has either been from the perspective of Linguistic scholars or has concentrated the focus of term evolution within a single domain of study. In this paper, we performed a case study to identify semantic evolution across different domains and identify examples of inter-domain semantic shifts. We initially used keywords as the basis of our search and executed an iterative process of following citations to find the initial mention of the concepts in the field. We found that a select set of keywords like ``semaphore'', ``polymorphism'', and ``ontology'' were mentioned within Computer Science literature and tracked the seminal study that borrowed those terms from original fields by citations. We marked these events as semantic evolution points. Through this manual investigation method, we can identify term evolution across different academic fields. This study reports our initial findings that will seed future automated and computational methods of incorporating concepts from additional academic fields.
\end{abstract}


\section{Introduction}
The evolution of new concepts in academia naturally occurs as new academic fields or streams evolve. When the new concept evolves, it is easier for scholars to use the existing keywords or concepts to explain and introduce the new one to the academic community. As academic domains expand, the concept within a domain can either be newly coined, change or drift, or decay. These changes can confuse researchers and hinder them from following those changes to build up further progress on those topics \cite{fokkens2016semantics}. However, even at the ontology level, it is difficult to maintain the quality of ontology changes or detect semantic changes within specific domains \cite{atkinson2021evolving}. Several studies have tracked semantic changes in linguistic perspectives \cite{beinborn2020semantic} and calculating semantic similarities \cite{pesquita2009semantic}, but tracking semantic evolutions across domains is yet to be a cultivated area. In this study, we conducted a preliminary case study on the semantic shift in certain concepts from one domain to another. Specifically, we chose concepts from Computer Science (CS) field and tried to identify the prior studies that mentioned the concepts first in the field. The reason why we chose CS to explore is that CS is considered a relatively recent academic field, and in order to explain the new concepts within the domains, the prior scholars must have used the already existing words or elaborations to explain their new ones. In that process, we postulate that various key concepts were borrowed from other academic fields. In that way, we would fill in the gap among academics that are closely and organically related and have influenced each other. The results of this study would highlight the unperceived connections between fields based on semantic shifts and borrowed terms.

\section{Background}
A semantic shift is a well-known and common phenomenon that can be detected in Linguistics. Words usage over a long period of time can be changed \cite{vanhove2008polysemy}. This shift could occur due to sociocultural factors, linguistic factors, or communicative-formal factors, among other reasons \cite{grzega2007english}. A semantic shift may also occur due to translation of a term from one language to another \cite{vandevoorde2020semantic}. 

Due to the complexity of the shift, many researchers have attempted and proposed models that can detect the semantic shift with computational measures. There have been various studies focusing on detecting semantic shifts within one language or certain corpus. Fišer and Ljubešić \cite{fivser2019distributional} used a distributional model for detecting word meanings in contemporary lexicography. They applied this model in their case study of detecting semantic shifts in Slovene tweets automatically. Martinc et al. \cite{martinc2019leveraging} introduced a novel approach for diachronic semantic shift detection that utilizes contextual embeddings and generates time-specific word representations based on BERT embeddings. Their results demonstrated that the model performs well in a domain-specific context, without the need for domain adaptation on large corpora as well as successful detection of short-term yearly semantic shifts. Hamilton et al. \cite{hamilton2016diachronic} identified that historical data on semantic shifts are limited, challenging further development on that matter. They proposed a methodology for quantifying semantic change by assessing word embeddings by testing them on six historical corpora encompassing four languages and two centuries to detect documented historical changes. Akidah \cite{akidah2013phonological} investigated the phonological and semantic transformations that occur when Arabic words are borrowed into Kiswahili, the two languages belonging to different language families. The result revealed a significant presence of Arabic-origin words in Kiswahili in the form of Kiswahili lexicon resulting from semantic changes such as broadening, narrowing, and shift.

Some studies tried to clarify and identify the semantic changes between different domains. Itagaki et al. \cite{itagaki2006detecting} measured the similarity of words between the technical and general domains by employing a method that quantifies the overlap in their domain-specific semantic spaces. These semantic spaces were calculated by grouping words that share syntactic similarities, specifically occurring in the same syntactic context.

However, there is a research gap in tracking and detecting inter-domain semantic shifts at the concept level with temporal elements. When new concepts evolve within one academic domain, the researchers would use the existing concepts to explain the new one, which helps to make the new usage of the term more easily understood. In this case study, we manually traverse research articles and follow citations to mark the temporal point for when the terms shifted from one domain to another. Even though this is preliminary work and is limited to individual case studies, the results provide concrete evidence that semantic shifts can be detected across different domains.

\section{Methodology}
We conducted an article search to track the origin of certain terms used in different domains. We assumed that the field that emerged prior is the origin of the terms. We focused on word drift within academia, namely, on academic journals that mentioned the terms.

As a first step, we chose three different terms known to be used in two or more different academic fields. Second, we conducted a keyword search on the Web of Science. We then limited the Web of Science fields to those relating to CS-related topic categories. We then identified the earlier usage based on temporal sorting. Once the earliest usage was identified, we manually checked to verify the seminal usage in that specific field.

\begin{figure}[h]
    \includegraphics[width=1.0\linewidth]{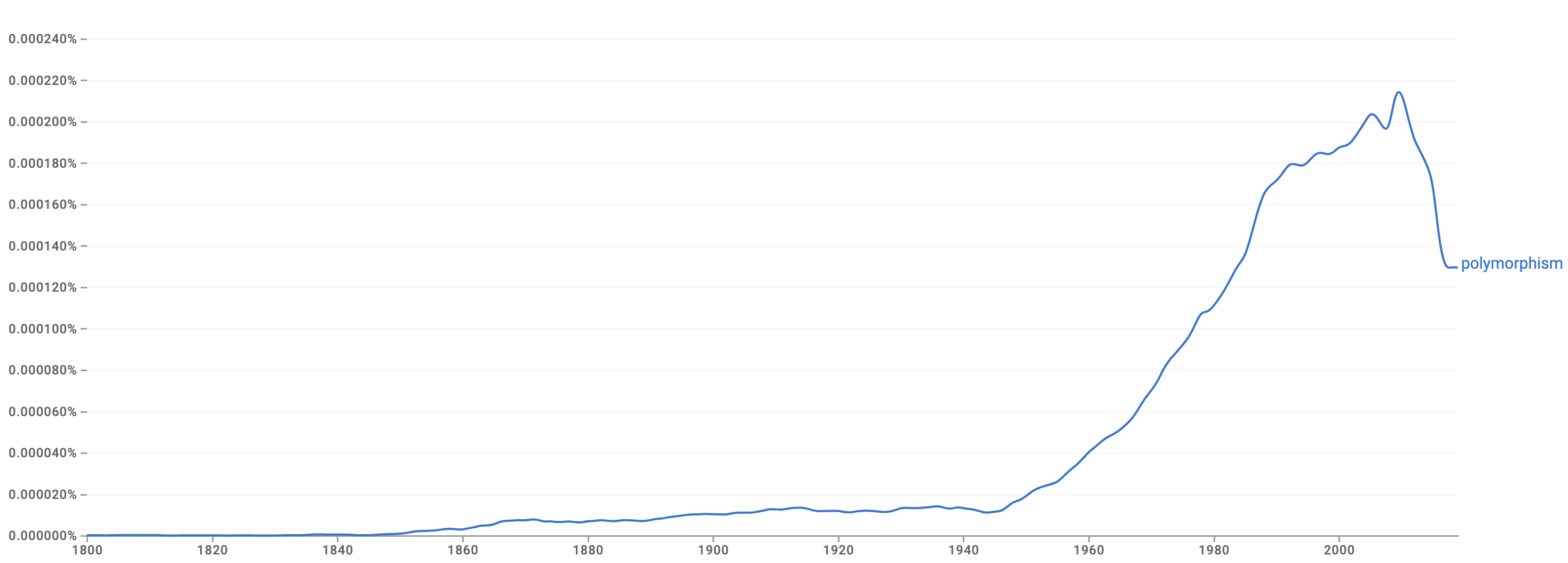}
    \caption{Gradual increase after the 1960s and steep peaks in 2000s in the usage of the term ``polymorphism'' as indicated by Google Ngram Viewer may reflect that the term has been used in different domains, possibly shifted in semantics.}
    \label{fig:ngram_polymorphism}
\end{figure}

\section{Case Study}

For this initial case study, we focused on three terms that have a well-recognized meaning in contemporary CS: polymorphism, semaphore, and ontology. We specifically investigated the CS field to choose concepts to explore because CS is considered as a relatively new domain compared to others. This would have more examples of borrowing terms from other fields when they explain and define their domain. After we chose three terms to search, we reviewed Google Ngram Viewer to see if identify any trend changes in usage that may be indicative of semantic shift. The graphs are presented in Figures ~\ref{fig:ngram_polymorphism}, ~\ref{fig:ngram_semaphore}, and ~\ref{fig:ngram_ontology}, respectively.

\subsection{Polymorphism}

\textit{Polymorphism} is a biological term that refers to the presence of two or more variant forms of a specific DNA sequence that can occur among different individuals or populations \cite{genome2019institute}. The term is also known in Computer Science (CS) as a single symbol representing various types \cite{cardelli1985understanding}. Since CS is a field that evolved later, we hypothesized that the term polymorphism was borrowed from Biology. We conducted the above search steps from the WoS to find the first CS article that mentioned polymorphism. The result showed that Coppo \cite{coppo1983semantics} is the earliest article that was captured within the database. When the paper was manually reviewed, Coppo cited an earlier article about the keyword by Milner \cite{milner1978theory}. Milner cited Strachey \cite{strachey1967fundamental} as the first mention of polymorphism. However, Strachey's article was not able to access in full text. This reflects well in Figure~\ref{fig:ngram_polymorphism} that polymorphism has been used frequently again after the 1980s.

After identifying the possible initial citation of polymorphism from CS, we tried to find when the citation of it stopped, as we hypothesized that when going backward in time, term usage without citation can reflect that the concept was accepted and prevalent in academic communities. We performed the same search condition and checked the papers from the most recent ones whether when the concept was mentioned the authors added citations to define or explain its meaning. We found that Houben \cite{houben2014fostering} explained polymorphism with citations within the study.

\subsection{Semaphore}

\textit{Semaphore} is a visual signaling method, usually by means of flags or lights, used to communicate between distant points \cite{britannica_2019}. In CS, a semaphore represents a protected variable or an abstract data type that restricts the access to shared resources in the programming environment \cite{anand2011context}. When it was searched on WoS, Denning et al. \cite{denning1981low} and Rony \cite{rony1981interfacing} were identified as the earliest articles. Rony \cite{rony1981interfacing} was unable to retrieve the full text. Also, when we looked through the study by Denning et al. \cite{denning1981low}, it did not include any citations or definitions that could lead to prior works. However, because the authors did not explain what semaphore means within the article, we assumed that the term was pervasive in that area.

\begin{figure}[h]
    \includegraphics[width=1.0\linewidth]{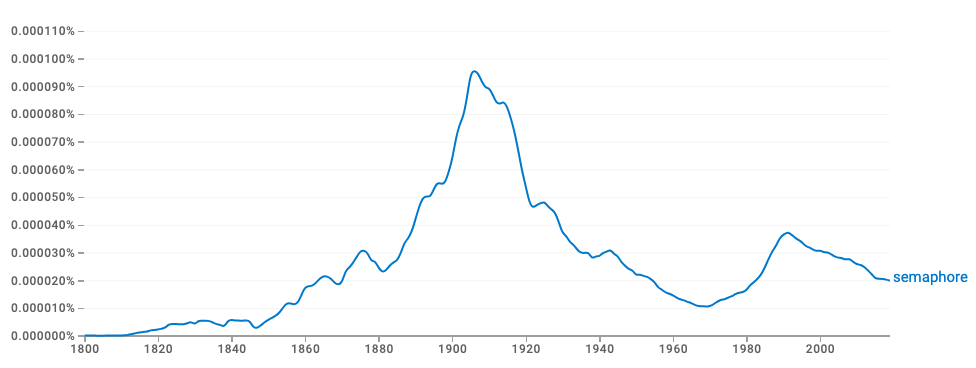}
    \caption{Multiple peaks in the usage of the term ``semaphore'' as indicated by Google Ngram Viewer may be indicative of the reuse of a term in another domain or the resurgence of a term with potentially different semantics.}
    \label{fig:ngram_semaphore}
\end{figure}

We did the same search and manually checked articles for citations from recent studies. We found a study from 2015 by Cicirelli et al. \cite{cicirelli2015approach} used citation in defining semaphores. Also, the citations were mentioned more frequently in the 1980s matches with the steep escalation of the Ngram graph from Figure~\ref{fig:ngram_semaphore}.

\subsection{Ontology}

\begin{figure}[h]
    \includegraphics[width=1.0\linewidth]{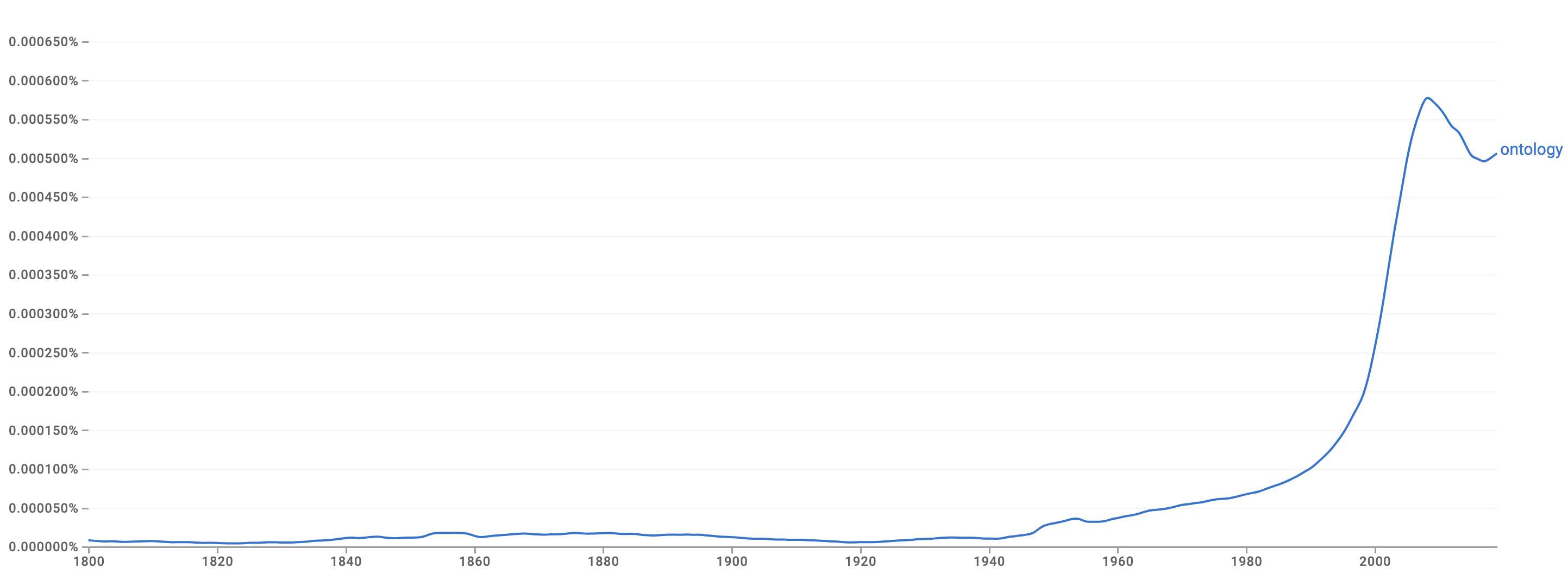}
    \caption{A high peak in the 2000s in the usage of the term ``ontology'' as indicated by Google Ngram Viewer may indicate that the term has been actively used in different domains with the possibility of semantic changes or shifts.}
    \label{fig:ngram_ontology}
\end{figure}

\textit{Ontology} is the philosophical study of \textit{being} in general or what applies naturally to everything that is real \cite{simons_2015}. Aristotle is often credited for introducing the concept in Book IV of his Metaphysics \cite{aristotle1933metaphysics}. In CS, ontology refers to ``an explicit specification of a conceptualization'' \cite{gruber1993translation}. When we searched in the database, Lenat and Guha's article \cite{lenat1990cyc} was identified as the earliest work within the WoS. We reviewed the full text was manually and found a citation of ontology within it: Quine \cite{quine1969ontological}. However, Quine's work cannot be retrieved for full text. 
When we searched to frame the timeline of when the academic studies stopped citing to define what ontology means, the concept represented a different pattern compared with the other two concepts. We found that authors tend to define the concept in their studies up until very recently, as shown in Figure~\ref{fig:ngram_ontology}, which has high picks after the 2000s. For instance, CB et al. (2023) quoted Ciotti and Tomasi's work from 2016 \cite{cb2023ontology,ciotti2016formal}. Park and Storey also cited several studies to explain what ontology means within their study in 2023 \cite{park2023emotion}.

\section{Discussion}

Semantic changes and evolution frequently occur as language changes over generations. This also occurs in terms and concepts within academia. As different domains evolve, each domain borrows terms or changes the semantics of the same terms to describe their new concepts or ideas. In that process, inter-domain semantic shifts take place. Tracking the shifts can reduce the ambiguity of terms and allows researchers from different domains to have more accessibility to other domain knowledge for new approaches to their studies.

In this case study, we identified when certain terms started to be used in one domain after having recognized usage and semantic in another prior. We limited the study to three terms that have different origins respectively and are known to be also used in the CS field. Since we used the academic search engine, we used citations to track the shifts within the academic communities and assumed that the communities would have accepted widely when authors stop using citations to describe or define them.

Based on the results, we could mark the approximate timeline for when the shifts were made. The interesting point was that the term ``ontology'' still includes citations in 2023 studies which resembles the Ngram graph that ontology has relatively a recent peak compared to the other two terms.

Our preliminary case study has some limitations including the limited time range of the corpus and manual search for the findings. Despite being in its early stages and confined to individual case studies, this preliminary research furnishes tangible proof that semantic shifts can be identified across various domains. Future works will include more computational methods including word embeddings, data crawling, and detecting semantic shifts with various language models. This may be used to identify the semantic shifts by including more domains and terms to present more dynamic changes over time by including larger sets of data.

\end{document}